\documentclass[aps,pra,preprint]{revtex4-2}

\usepackage{graphicx}
\usepackage{dcolumn}
\usepackage{bm}
\usepackage{amsmath}
\usepackage{siunitx}
\usepackage{gensymb}
\usepackage{chemfig}

\begin{document}

\title{Magnetostriction and Temperature Dependent Gilbert Damping in Boron Doped Fe$_{80}$Ga$_{20}$ Thin Films}

\author{Zhixin Zhang}
\email{zhixin2@illinois.edu}
\affiliation{%
Department of Materials Science and Engineering and Materials Research Laboratory, The Grainger College of Engineering, University of Illinois Urbana-Champaign, Urbana, Illinois, 61801, USA
}%

\author{Jinho Lim}
\affiliation{%
Department of Materials Science and Engineering and Materials Research Laboratory, The Grainger College of Engineering, University of Illinois Urbana-Champaign, Urbana, Illinois, 61801, USA
}%

\author{Haoyang Ni}
\affiliation{%
Department of Materials Science and Engineering and Materials Research Laboratory, The Grainger College of Engineering, University of Illinois Urbana-Champaign, Urbana, Illinois, 61801, USA
}%

\author{Jian-Min Zuo}
\affiliation{%
Department of Materials Science and Engineering and Materials Research Laboratory, The Grainger College of Engineering, University of Illinois Urbana-Champaign, Urbana, Illinois, 61801, USA
}%

\author{Axel Hoffmann}
\email{axelh@illinois.edu}
\affiliation{%
Department of Materials Science and Engineering and Materials Research Laboratory, The Grainger College of Engineering, University of Illinois Urbana-Champaign, Urbana, Illinois, 61801, USA
}%

\begin{abstract}
Magnetic thin films with strong magnetoelastic coupling and low Gilbert damping are key materials for many magnetoelectric devices. Here, we investigated the effects of boron doping concentration on magnetostriction and temperature dependent Gilbert damping in magnetron sputtered (Fe$_{80}$Ga$_{20}$)$_{1-x}$B$_{x}$ films. A crystalline to amorphous structural transition was observed for a boron content near 8\% and coincided with a decrease in coercivity from 76~Oe to 3~Oe. A 10\% doping concentration is optimal for achieving both large magnetostriction of 48.8~ppm and low Gilbert damping of $6 \times 10^{-3}$. The temperature dependence of the damping shows an increase at low temperatures with a peak around 40~K and we associate the relative increase $\Delta\alpha/\alpha_{RT}$ with magnetoelastic contributions to the damping, which has a maximum of 55.7\% at 8\% boron. An increase in the inhomogeneous linewidth broadening was observed in the structural transition regime at about 8\% boron concentration. This study suggests that incorporation of glass forming elements, in this case boron, into Fe$_{80}$Ga$_{20}$ is a practical pathway for simultaneously achieving enhanced magnetoelastic coupling and reduced Gilbert damping.
\end{abstract}

\maketitle

\section{\label{sec:introduction} Introduction}
Magnetoelastic materials, which alter their magnetic properties under applied mechanical stress, have gained increasing interest due to their energy efficiency and versatility in applications such as actuators, sensors, and energy harvesters \cite{s20051532}. One type of application where magnetoelastic coupling can provide new functionality is given by surface acoustic waves (SAW) devices. SAW are utilized due to their short wavelengths on the micrometer scale at gigahertz frequencies. At the same time, magnons, which are the collective excitations in magnetic materials, have even smaller wavelengths in the nanometer range. 

SAWs are widely used in today’s world for communication technologies \cite{7892034, 6332680}, biosensing \cite{ji2020aptamer, BISOFFI20081397}, and signal processing \cite{Katzman:22}. 
For these applications, SAW devices use interdigitated transducers to efficiently excite or detect coherent SAWs on piezoelectric substrates \cite{10.1063/1.1787897, 10.1063/1.5142673, 10.1063/1.1654476}. 
In addition, propagating SAWs can couple to magnons through magnetoelastic coupling \cite{10.1063/5.0157520}, realizing nonreciprocal SAW which is fundamental for miniaturized on-chip acoustic isolators and circulators \cite{shao2020non, doi:10.1126/sciadv.abc5648, PhysRevApplied.12.054061, https://doi.org/10.1002/aelm.202100263}. Beyond the classical regime, SAWs have also been explored in the quantum regime for single quantum coherent operations \cite{satzinger2018quantum, dumur2021quantum,PhysRevLett.119.180505,doi:10.1126/science.aaw8415}. Towards this end, nonreciprocal SAWs can be useful for microwave based quantum communication and information processing, mitigating noise and establish non-reciprocal state-transfer for superconducting quantum devices \cite{10.1063/5.0157520}. 

Nonreciprocal SAW transmission resulting from magnon-phonon coupling in the classical regime has been previously observed in various configurations such as bilayer structures composed of FeGaB/Al$_{2}$O$_{3}$/FeGaB \cite{PhysRevApplied.18.044003, doi:10.1126/sciadv.abc5648}, synthetic antiferromagnets with CoFeB/Ru/CoFeB \cite{doi:10.1021/acsaelm.3c01709, doi:10.1021/acsaelm.3c00844}, anti-magnetostrictive bilayers of Ni/Ti/FeCoSiB \cite{10.1063/5.0196523}, and layered antiferromagnets of CrCl$_{3}$ \cite{PhysRevLett.131.196701}. Among the materials capable of generating large SAW nonreciprocity, FeGaB is a strong candidate due to its large magnetostriction. At the same time, FeGaB has low Gilbert damping, which was suggested to be useful for observing large SAW nonreciprocity \cite{PhysRevApplied.13.044018, 10.1063/5.0196523}. Although previous study suggested boron doping can increase the magnetostriction and decrease FMR linewidth for Fe$_{1-x}$Ga$_x$ with $x$ varying from 9 to 17 \cite{10.1063/1.2804123}, studies still remain scarce on boron concentration dependent magnetostriction and magnetic damping for the specific Fe$_{80}$Ga$_{20}$ stoichiometry ratio, especially at cryogenic temperatures which are essential for quantum operations.

In this work, magnetostriction and temperature dependent Gilbert damping in magnetron sputtered (Fe$_{80}$Ga$_{20}$)$_{1-x}$B$_{x}$ films were systematically investigated. X-ray diffraction (XRD) revealed a suppressed diffraction peaks in high boron doping regime, consistent with the vanishing of the lattice fringes observed in high-resolution electron microscopy (HRTEM), indicating a transition from polycrystalline to amorphous. Significant enhancement of magnetostriction was observed in the amorphous regimes, with the magnetostriction being maximized at 10\% boron. Concomitantly, the Gilbert damping was reduced in the amorphous regime. Gilbert damping at cryogenic temperatures was measured along with a newly introduced parameter $\Delta\alpha/\alpha_{RT}$ for quantifying the observed peak in the damping at around 40~K. It is noticed that $\Delta\alpha/\alpha_{RT}$ followed a similar trend as the magnetostriction, indicating potential magnetoelastic coupling contribution to the damping peak at 40~K. 

\section{\label{sec:level2} Experimental Details}
\begin{figure}[t]
        \centering
        \includegraphics[width=0.7\columnwidth]{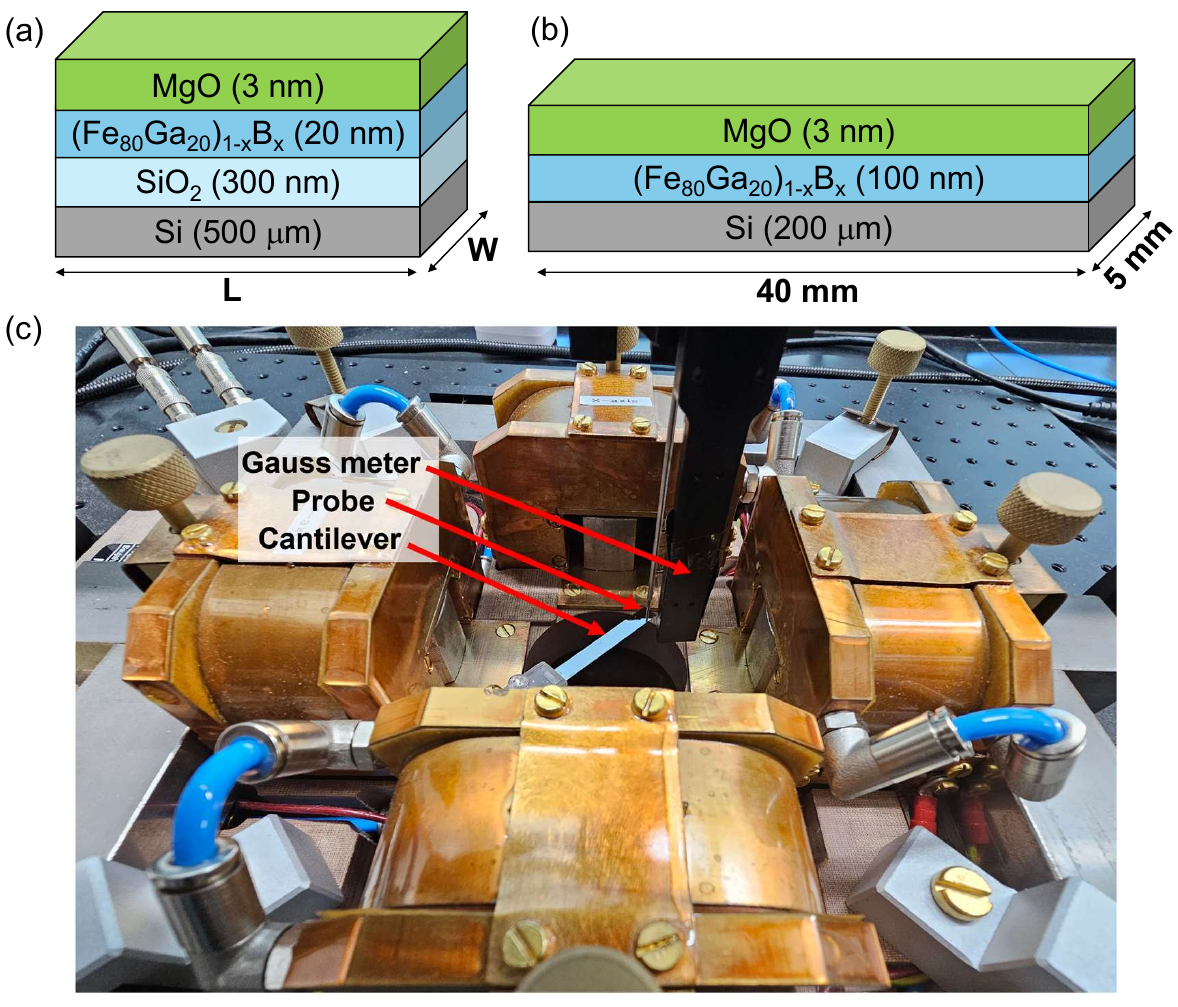}
        \caption{(a) Sample geometry and layer information for 20-nm FeGaB films where L $\times$ W = 6~mm $\times$ 5~mm and 4~mm $\times$ 4~mm and (b) geometry of the cantilever with a 100-nm FeGaB film for magnetostriction measurement. (c) Magnetostriction measurement setup consisting of two pairs of coils for generating an in-plane rotating field, with the optical probe tip detecting vibration at the free end of the cantilever.}
        \label{setup}
\end{figure}

(Fe$_{80}$Ga$_{20}$)$_{1-x}$B$_{x}$ ($x$ = 2 to 16\%) films of thickness 20~nm on 500-$\mu$m Si substrates with lateral dimensions of 5~mm $\times$ 6~mm and 4~mm $\times$ 4~mm as shown in Fig.~\ref{setup}(a) were synthesized by {\em dc} magnetron sputtering at room temperature without heat treatment. The base pressure was less than $3 \times 10^{-8}$ Torr and an Ar pressure of 6~mTorr was used for the deposition. To gain precise control on boron doping concentration, three sputter targets, Fe, Fe$_{50}$Ga$_{50}$, and Fe$_{60}$B$_{40}$, were sputtered simultaneously. Linear fittings of growth rates calibrated by X-ray reflectivity (XRR) as a function of individual sputter gun power were used to calculate the power required for a specific nominal boron concentration. A 3-nm MgO capping layer was deposited to prevent oxidation. For magnetostriction measurements, (Fe$_{80}$Ga$_{20}$)$_{1-x}$B$_{x}$ films of thickness 100~nm were deposited on 200-$\mu$m thick Si cantilever with lateral dimensions of 5~mm $\times$ 40~mm, as indicated in Fig.~\ref{setup}(b).  

TEM specimens were prepared by depositing 16 nm (Fe$_{80}$Ga$_{20}$)$_{100}$B$_{0}$ and (Fe$_{80}$Ga$_{20}$)$_{84}$B$_{16}$ thin film on 8 nm SiO$_{2}$ TEM membrane grid (Ted Pella, Inc.) using the same deposition conditions as mentioned above. Prior to deposition, membrane grids were in-situ plasma cleaned to remove possible surface contamination. High-resolution transmission electron microscopy (HRTEM) and selected-area electron diffraction (SAED) were performed using a Thermo Fisher Talos F200X, operated at 200 kV.

Rutherford Backscattering (RBS) was used to verify the Fe to Ga ratio. XRR was used to measure the thickness of (Fe$_{80}$Ga$_{20}$)$_{1-x}$B$_{x}$ films by fitting the data with the GenX software. Structural characterization by XRD was performed on Bruker D8 Advance using the cantilever samples. Magnetic hysteresis loops were measured on a Quantum Design superconducting quantum interference device (SQUID) magnetometer. 

Magnetostriction was measured by a cantilever deflection technique similar to Ref.\cite{10.1063/1.5065486} using a custom built magnetostriction tester shown in Fig.~\ref{setup}(c). Two pairs of coils driven by two 90-degree out of phase low frequency AC currents produced an in-plane rotating magnetic field. In the field center, the cantilever coated with (Fe$_{80}$Ga$_{20}$)$_{1-x}$B$_{x}$ was clamped on one end and the vibrational deflection of the free end was measured by a light reflectance based MTI-2100 Fotonic sensor with 1-nm displacement resolution. The field rotation frequency was set to 16.3~Hz to avoid mechanical resonance of the setup. To extract small amplitude vibrational signal, the voltage output from the Fotonic sensor, which is linearly related to the detected displacement, was measured with a SR-830 lock-in amplifier at the second harmonic cantilever vibrational frequency of 32.6~Hz. Background subtraction was performed by measuring the response of a bare Si cantilever to account for potential probe vibrations. The magnetic field strength and homogeneity around the cantilever were measured and verified by a Senis Gaussmeter. 

Dynamic magnetic properties were investigated using field modulation ferromagnetic resonance (FMR) in a cryostat capable of maintaining sample temperature from 2~K to 300~K. The FMR setup involved a {\em rf} signal generator, a coplanar waveguide coupled to the sample through a flip-chip mounting, a diode, and a lock-in amplifier referenced to the field modulation signal with frequency 17~Hz. The external magnetic field and the modulation field were both applied in plane parallel to the signal line of the coplanar waveguide. Resonance linewidth fitting algorithms were incorporated into the measurement process for automatically adjusting modulation field strength and field sweep rate to achieve large signal to noise ratio (SNR) without inducing artificial peak broadening. 

\section{\label{sec:level3} Results}
\subsection{Structural Characterization}
The structures for the Fe$_{80}$Ga$_{20}$ films with various boron doping concentrations were measured by XRD and the spectra is shown in Fig.~\ref{XRD}. The full-range scans containing Si substrate peaks can be found in Fig. S1 in the supplementary materials. Two peaks associated with our films were identified, with one major (110) peak at 2$\theta$ = 44\degree\ and a minor (211) peak at 2$\theta$ = 81.5\degree\ that vanished for films with more than 4\% boron doping. The inset of Fig.~\ref{XRD} shows the background corrected spectra around the (110) peak. The peak intensity was roughly the same for the Fe$_{80}$Ga$_{20}$ without boron doping and Fe$_{80}$Ga$_{20}$ with 4\% boron doping. When boron concentration reached 6\%, a slightly lower peak was noticed, indicating reduced crystallinity. With 12\% boron, the formation of amorphous phase at the heavily doped regime was signified by a broad hump. Between 6\% to 10\% boron, a mixed phase regime was likely to exist. This XRD result was further confirmed by selected-area electron diffraction (SAED), shown in Fig. S2 of the supplementary materials.

\begin{figure}[htb]
        \centering
        \includegraphics[width=0.7\columnwidth]{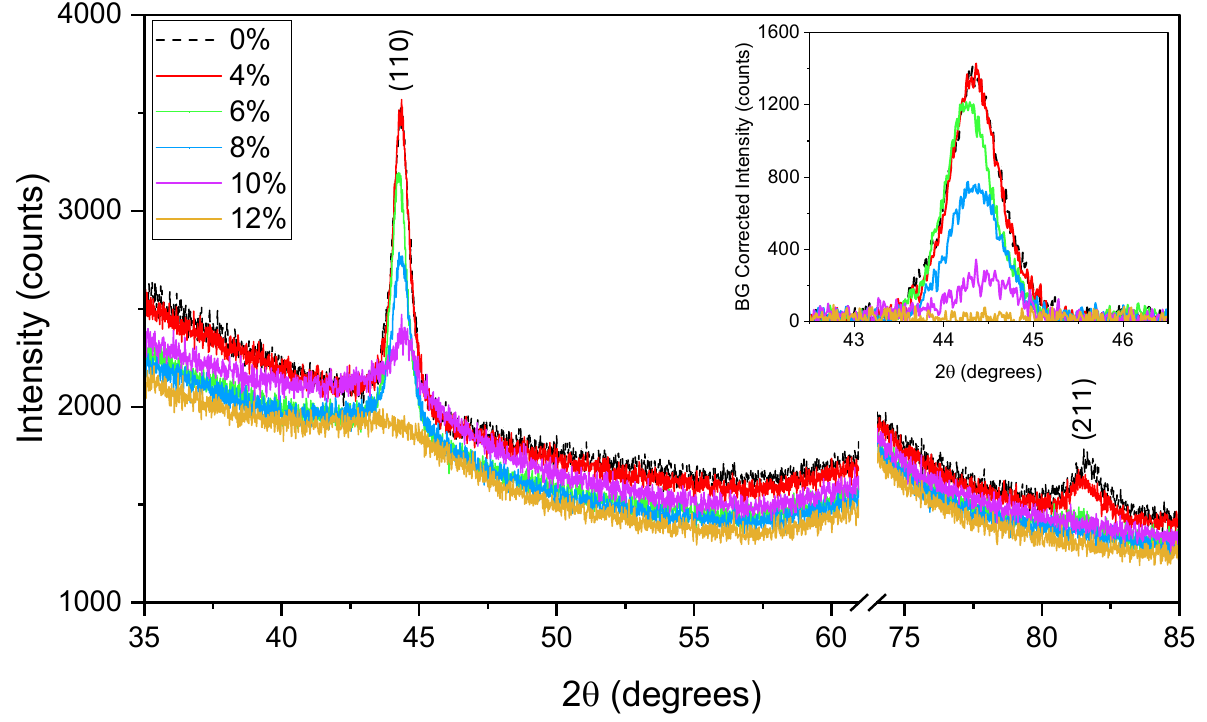}
        \caption{XRD spectra for undoped Fe$_{80}$Ga$_{20}$ film and boron doped (Fe$_{80}$Ga$_{20}$)$_{1-x}$B$_{x}$ films with $x = 4$, 6, 8,  10, and 12\%. The inset shows background corrected spectra around (110) peak.}
        \label{XRD}
\end{figure}

\begin{figure}[t]
        \centering
        \includegraphics[width=0.7\columnwidth]{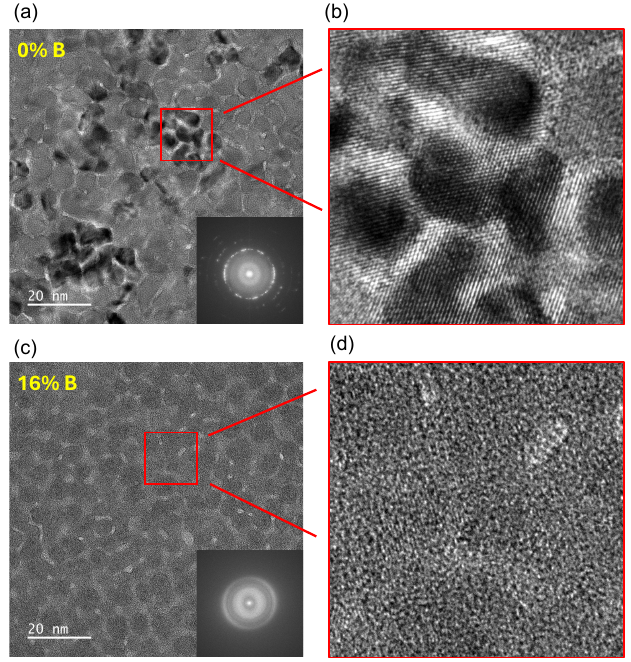}
        \caption{HRTEM images of (Fe$_{80}$Ga$_{20}$)$_{100}$B$_{0}$ (a, b) and (Fe$_{80}$Ga$_{20}$)$_{84}$B$_{16}$ (c, d). (b, d) are enlarged image from the regions labeled by red box in (a, c). Insets of (a, c) are the FFTs of the HRTEM images of (Fe$_{80}$Ga$_{20}$)$_{100}$B$_{0}$ and (Fe$_{80}$Ga$_{20}$)$_{84}$B$_{16}$. }
        \label{TEM}
\end{figure}

Fig.~\ref{TEM} shows HRTEM images of (Fe$_{80}$Ga$_{20}$)$_{100}$B$_{0}$ and (Fe$_{80}$Ga$_{20}$)$_{84}$B$_{16}$. In the case of non-doped Fe$_{80}$Ga$_{20}$, clear lattice fringes can be seen with random orientation showing an overall nanocrystalline nature of the thin film. In contrast, no clear lattice fringes can be observed in Fe$_{80}$Ga$_{20}$ with 16\% boron doping, indicating its overall amorphous structure.

\subsection{Static Magnetic Properties}
\begin{figure}[htb]
        \centering
        \includegraphics[width=0.9\columnwidth]{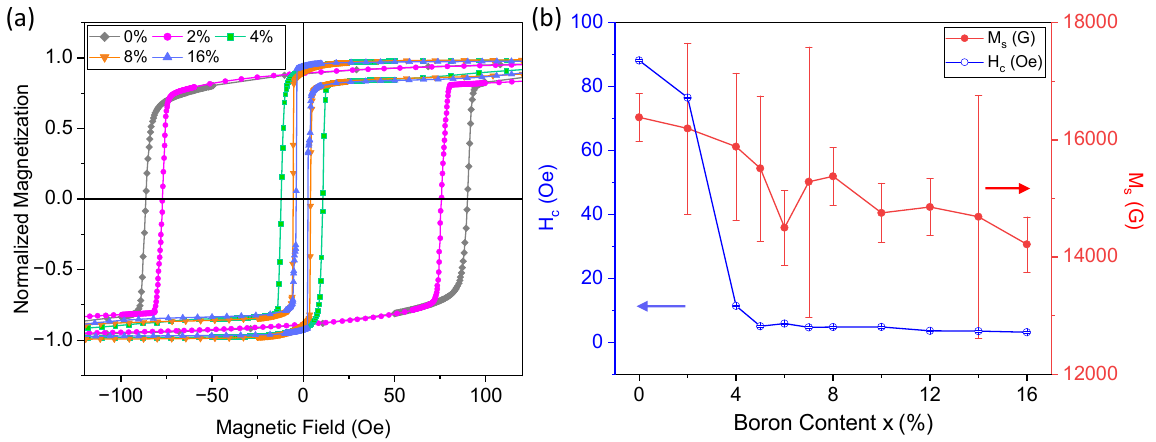}
        \caption{(a) Magnetic hysteresis loops of undoped Fe$_{80}$Ga$_{20}$ and (Fe$_{80}$Ga$_{20}$)$_{1-x}$B$_{x}$ films with $x = 2$, 4, 8, and 16\% measured along in-plane direction at room temperature and (b) the corresponding coercivity and saturation magnetization obtained from the hysteresis loops and magnetization curves.}
        \label{MPMS}
\end{figure}

Static magnetic properties were investigated by magnetic hysteresis loop measurements with an in-plane magnetic field. Fig.~\ref{MPMS}(a) shows the magnetization hysteresis loops for the undoped Fe$_{80}$Ga$_{20}$ film and 2, 4, 8, and 16\% boron doped Fe$_{80}$Ga$_{20}$ films. A rapid decrease in coercivity from around 80~Oe to 10~Oe was observed when increasing boron doping concentration from 2\% to 4\%, while such reduction was less pronounced upon further increasing boron concentration beyond 8\%. This can be explained by the fact that the films reached the phase boundary and started to form the amorphous phase with consistently low coercivity. To extract saturation magnetization, large in-plane magnetic field up to 4~T was applied and the magnetization curves are shown in Fig. S3 in the supplementary material. In Fig.~\ref{MPMS}(b), the coercivity and saturation magnetization are plotted against the boron doping concentration.  We found both coercivity and saturation magnetization decreased with boron concentration. The decrease in saturation magnetization is in agreement with the reduction of the density of magnetic atoms and has a roughly linear trend that is similar to other well-studied doped magnetic systems such as Co-Fe-C \cite{PhysRevApplied.12.034011}. The reduction in coercivity is consistent with the structural transition from crystalline to amorphous.

\subsection{Magnetostrictive Property}
The custom built magnetostriction measurement system was used to investigate the magnetostrictive properties of (Fe$_{80}$Ga$_{20}$)$_{1-x}$B$_{x}$ coated on thin Si cantilevers. Due to the induced
strain in the magnetic layer bonded to the cantilever, the cantilever developed curvature and thus a vertical displacement at the free end. With the rotating field, the strain state in the magnetostrictive film was constantly changing, leading to a detectable vibration at the free end. The magnetostriction constant can then be derived from the measured deflection using Eq.~\ref{Eq:magnetostriction} \cite{1059251},

\begin{eqnarray}
\label{Eq:magnetostriction}
    d &=& \lambda\dfrac{3t_fl^2E_f(1-v_s)}{t_s^2E_s(1+v_f)},
\end{eqnarray}

where $t_{f}$ and $t_{s}$ are the thicknesses of the film and the substrate, respectively, l is the distance between the clamped position of the cantilever to the probe location. $E_{f}$ and $E_{s}$ are the Young’s modulus, and $v_{s}$ and $v_{f}$ are the Poisson’s ratios of the film and the substrate, respectively. Here, $t_{f}$ was set to the actual film thickness measured by XRR, $E_{s}$ and $v_{s}$ of the [110] Si cantilever was set to 169~GPa and 0.064, respectively \cite{5430873}. Following the convention, the Young’s modulus and Poisson ratio of FeGaB films were approximated to be \(\frac{E_f}{1+v_f}\) = 50~GPa \cite{10.1063/1.2980034}. To check the accuracy of our setup, materials with previously reported magnetostriction, such as permalloy (zero magnetostriction), Co, FeGaB, Co$_{40}$Fe$_{60}$ and Co$_{25}$Fe$_{75}$ were synthesized using magnetron sputtering and the measured magnetostriction values were compared with the literature reported values, as shown in Fig. S4 in the supplementary material.

Fig.~\ref{Magnetostriction}(a) shows the measured magnetostriction as a function of applied rotating magnetic field strength for the 100-nm Fe$_{80}$Ga$_{20}$ with varying boron doping concentrations. With increasing field, the magnetostriction response became stronger, until reaching the maximum magnetostriction at the saturation field after which it plateaued. In Fig.~\ref{Magnetostriction}(b), the saturation magnetostriction constants, defined as the maximum point on the magnetostriction curve, were plotted as a function of boron concentrations. We noticed the saturation magnetostriction significantly enhanced from ${7.9\pm0.6}$~ppm for 4\% boron to ${48.8\pm0.9}$~ppm for 10\% boron, representing more than six times improvement. In addition, we noticed the saturation magnetostriction started to decrease when doping past 10\% boron, away from the mixed-phase boundary regime. This is similar to other promising magnetoelastic materials that have largest magnetostriction in the mixed-phase boundary regimes \cite{90291b70510841818062f2edfc3f517f, PhysRevApplied.12.034011, 10.1063/1.2804123}, which is 8\% to 10\% boron in our case.

\begin{figure}[t]
        \centering
        \includegraphics[width=0.9\columnwidth]{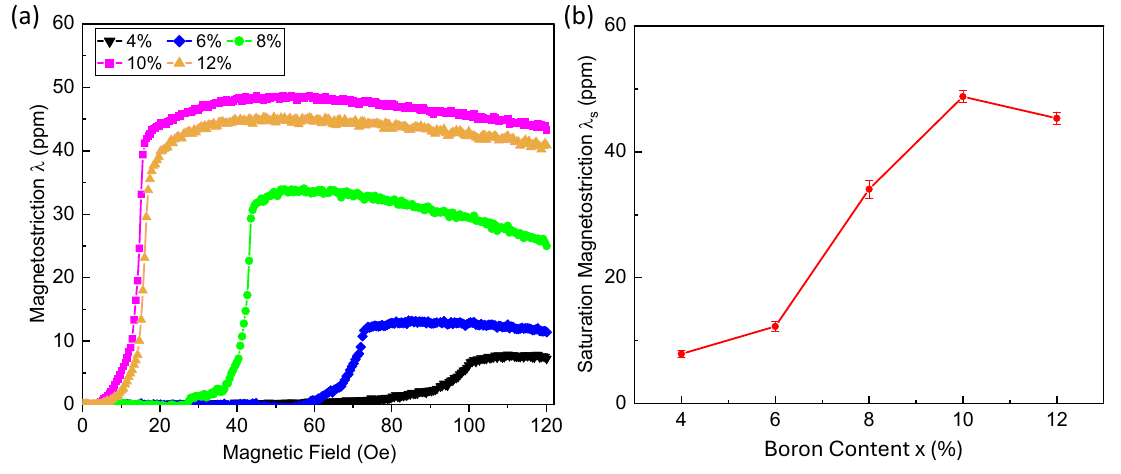}
        \caption{(a) Magnetostriction constants as a function of applied in-plane field strength of 80-nm (Fe$_{80}$Ga$_{20}$)$_{1-x}$B$_{x}$ films with $x = 4$, 6, 8, 10, 12\%
        measured using cantilever deflection. (b) Saturation magnetostriction as a function of boron concentration.}
        \label{Magnetostriction}
\end{figure}

\subsection{Dynamic magnetization and temperature dependent Gilbert damping}
\begin{figure}[htb]
        \centering
        \includegraphics[width=0.9\columnwidth]{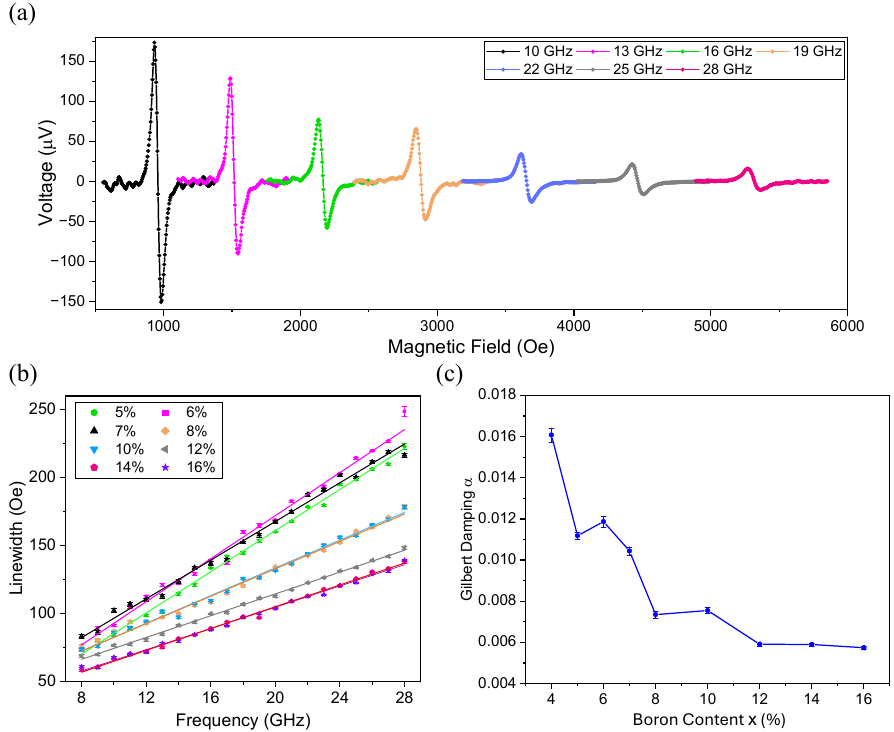}
        \caption{(a) FMR spectrum at different excitation frequencies for 20-nm (Fe$_{80}$Ga$_{20}$)$_{88}$B$_{12}$ film measured at room temperature.  (b) Linear fitting of resonance linewidth vs. frequency for Fe$_{80}$Ga$_{20}$ with different boron doping concentrations. (c) Extracted Gilbert damping as a function of boron content in boron doped Fe$_{80}$Ga$_{20}$ films at room temperature.}
        \label{RTDamping}
\end{figure}

Magnetization dynamics were investigated by field modulation in-plane FMR. Fig.~\ref{RTDamping}(a) shows the voltage detected from the lock-in amplifier as a function of applied magnetic field for the 20-nm Fe$_{80}$Ga$_{20}$ film with 12\% boron doping. Due to the presence of a small modulation field, a derivative Lorentzian peak was observed. Resonance field $H_{0}$ and linewidth ${\Delta H}$ were extracted by fitting the derivative of Lorentzian peak to Eq.~\ref{Eq:peakfitting},

\begin{eqnarray}
\label{Eq:peakfitting}
    V &=& S\dfrac{4\Delta H(H-H_0)}{[4(H-H_0)^2+\Delta H^2]^2} \nonumber\\
    &&+ A\dfrac{\Delta H^2-4(H-H_0)^2}{[4(H-H_0)^2+\Delta H^2]^2},
\end{eqnarray}

where ${S}$ and ${A}$ are the coefficients of symmetric and anti-symmetric components, ${\Delta H}$ is the full width at half maximum, ${H}$ is the external applied field, and $H_0$ is the resonance field. In Fig.~\ref{RTDamping}(b), the resonance linewidth was plotted against frequency for Fe$_{80}$Ga$_{20}$ with different  boron doping concentrations and Gilbert damping was extracted based on Eq.~\ref{Eq:dampingfitting},

\begin{eqnarray}
\label{Eq:dampingfitting}
    \Delta H &=& \dfrac{4\pi}{\gamma}\alpha f + \Delta H_0~,
\end{eqnarray}

where ${\Delta H}$ is the full width at half maximum, ${\gamma}$ is the gyromagnetic ratio, ${f}$ is the applied microwave frequency, and ${\Delta H_0}$ is the inhomogeneous linewidth broadening. The gyromagnetic ratio is calculated from the fitted electron g-factor of 2.11 which has negligible variation with boron concentration or temperature as shown in Fig. S5 in the supplementary material. Fig.~\ref{RTDamping}(c) shows the extracted Gilbert damping as a function of boron doping concentration. We noticed the damping significantly reduced from $1.6 \times 10^{-2}$ with 4\% boron to $5.7 \times 10^{-3}$ with 16\% boron doping, representing more than two-fold damping suppression. In addition, similar to the magnetostriction trend, the damping trend is also divided into two regimes: high damping regime where the film is polycrystalline and low damping regime where the film is amorphous. 

\begin{figure}[htb]
        \centering
        \includegraphics[width=0.9\columnwidth]{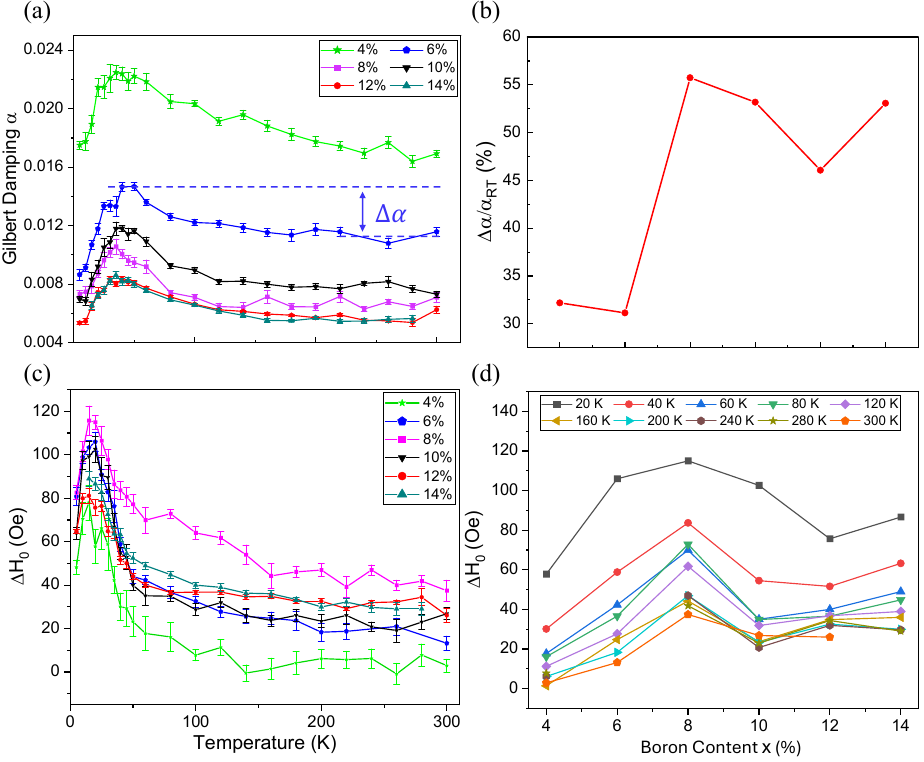}
        \caption{(a) Gilbert damping as a function of temperature for 20-nm (Fe$_{80}$Ga$_{20}$)$_{1-x}$B$_{x}$ films with $x$~= 4, 6, 8, 10, 12, and 14\%. (b) Relative peak intensity $\Delta\alpha/\alpha_{RT}$ as a function of boron content~$x$. (c, d) Inhomogeneous linewidth broadening as a function of (c) temperature and (d) boron content~$x$.}
        \label{LowTDamping}
\end{figure}

To investigate the temperature dependent Gilbert damping, the sample and the coplanar waveguide were mounted in a cryostat with variable temperature down to 2 K. The temperature dependent Gilbert dampings for different boron doped Fe$_{80}$Ga$_{20}$ were measured and plotted in Fig.~\ref{LowTDamping}(a). Interestingly, the Gilbert damping did not show a monotonic behavior with decreasing temperature. Instead, the damping first increased with decreasing temperature, reaching a maximum at around 40~K, and then rapidly decreased to its minimum at 5~K. Although this damping peak existed in all samples, the peak intensity varied depending boron doping concentrations. Here, to facilitate discussion, we define the relative damping increase $\Delta\alpha/\alpha_{RT}$ as the ratio of the damping increase $\Delta\alpha$ at the 40-K peak ($\alpha_{Max}-\alpha_{RT}$) to the average value of damping $\alpha_{RT}$ around room temperature (260~K, 280~K, and 300~K). $\Delta\alpha/\alpha_{RT}$ was calculated and plotted in Fig.~\ref{LowTDamping}(b). We observed that the relative damping increase $\Delta\alpha/\alpha_{RT}$ was higher in the amorphous regime with high magnetoelastic coupling compared to the crystalline regime where magnetoelastic coupling was small. Furthermore, we noticed the damping at 5~K was always lower than the room temperature damping regardless of boron doping level, which is promising for quantum devices operating at mK temperatures. 

In addition to the Gilbert damping, the inhomogeneous linewidth broadening ${\Delta H}$, which is associated with interface magnetic non-uniformity or defects, was extracted using on Eq.~\ref{Eq:dampingfitting} and plotted as a function of temperature and boron concentration in Fig.~\ref{LowTDamping}(c) and (d), respectively. As the temperature decreased from 300 K, the inhomogeneous linewidth broadening increased to a peak at around 20 K, after which it decreased until 5 K. The increase in inhomogeneous linewidth broadening is likely attributed to the interfacial strain caused by different thermal expansion coefficient of the film and the substrate. However, the origin of the peaks at around 20 K is not clear and requires further investigation. Boron doping also affects the inhomogeneous linewidth broadening in a way that it first increased with boron doping until reaching a peak at 8\% and then decreased as more boron was incorporated, consistent with boron's influence on the film crystallinity. At low boron doping level, the film is crystalline which is in favor of more uniform local magnetic moments. At the mix-phased regime around 8\%, magnetic domains have more randomness, leading to larger inhomogeneous linewidth broadening. Further boron doping brings the film into the amorphous phase with enhanced homogeneity. 

\section{\label{sec:level4} Discussion}
We have observed for the (Fe$_{80}$Ga$_{20}$)$_{1-x}$B$_{x}$ films a strong boron concentration dependence on coercivity, magnetostriction, Gilbert damping, and in the case of temperature dependent studies, the Gilbert damping peak intensity. The concentration dependence here can be classified into two regimes based on doping levels: a low boron doping regime with less than 8\% boron where the magnetostriction is low, damping is high but with a small damping peak enhancement around 40 K, and a high boron doping regime with boron greater than 8\% where we noticed improved saturation magnetostriction, suppressed Gilbert damping, and a stronger Gilbert damping enhancement around 40~K. As a metallic alloy, FeGa undergoes phase transition at around 19\% Ga concentration, transitioning to an ordered ${D0_3}$ phase, where Ga atoms have preferential occupation sites along two perpendicular $\langle 110 \rangle$  directions, compared to the disordered A2 phase at low Ga concentration where Ga atoms are randomly distributed in the BCC lattice. The magnetostriction monotonically increases with increasing Ga content until 19\% after which it starts to decrease due to the formation of the ordered ${D0_3}$ phase \cite{10.1063/1.1540130}. However, in both ${D0_3}$ and A2 phases, the lattices are well defined. Here, by introducing small boron atoms into the lattice, the lattice was gradually eliminated, leading to improved magnetostriction and damping performance. 

A previous study \cite{PhysRevB.103.L220403} proposed that the Gilbert damping in magnetostrictive material can have contribution from magnetoelastic damping based on a phonon relaxation mechanism, which is consistent with our observations. In our study, the boron dependent magnetostriction was measured at room temperature. It is possible that the magnetostriction itself also has a temperature dependence, but previous study on bulk FeGa alloys with similar Ga concentration reported that the magnetoelastic coupling only increased by 15\% when decreasing from room temperature to 4~K \cite{10.1063/1.1856731}, indicating the magnetoelastic coupling itself has a weak temperature dependence. It is noted that in our study both the Gilbert damping and the inhomogeneous linewidth broadening showed a peak in their temperature dependence, but with different peak positions. In the case of Gilbert damping, the peak occurred around 40 K, while the peak for the inhomogeneous linewidth broadening was located at 20 K, indicating different underlying mechanisms.

To further correlate the damping peak at 40~K to magnetoelastic coupling, we compared the temperature dependent damping of (Fe$_{80}$Ga$_{20}$)$_{1-x}$B$_{x}$ to that of permalloy (Ni$_{81}$Fe$_{19}$), which is known to be non-magnetostrictive \cite{10.1063/1.328971}. Our magnetostriction measurement confirmed the lack of magnetostriction in the permalloy and temperature dependent Gilbert damping scan showed no peak at 40~K, as shown in Fig. S6 in the supplementary material. We note in the case of permalloy, that the temperature dependence of Gilbert damping is also related to the film thickness and a damping peak at 40~K was previously reported for permalloy with less than 10~nm thickness \cite{zhao2016experimental}. Nevertheless, the ratio of damping increases $\Delta\alpha/\alpha_{RT}$ in those samples are at most 13\%, much smaller than what we observed in magnetostrictive FeGaB. Therefore, the different peak height observed in FeGa with varying boron concentration is most likely related to magnetoelastic coupling. However, further experimental or theoretical investigations are needed to reveal the underlying mechanism of such Gilbert damping peak. 

\section{\label{sec:level5} Conclusion}
In summary, we alloyed different concentrations of boron into Fe$_{80}$Ga$_{20}$ thin films and studied their saturation magnetostriction, coercivity, and temperature dependent Gilbert damping. We observed a crystalline to amorphous phase transition with increasing boron concentration and identified a phase boundary between 8\% to 10\% boron concentration. The decrease in coercivity while incorporating more boron is consistent with the gradually reduced crystallinity. The amorphous regime has higher magnetostriction and lower Gilbert damping, which are both desired for generating large SAW nonreciprocity. Furthermore, We conducted temperature dependent Gilbert damping study and noticed a boron concentration dependent damping peak at around 40~K and this relative damping increase $\Delta\alpha/\alpha_{RT}$ has a boron concentration dependence consistent to that of the magnetostriction. Combined with the vanishing of the damping peak observed in nonmagnetostrictive permalloy thin film, this indicates the damping peak in the temperature scan could originate from magnetoelastic coupling. 

\section{\label{sec:level6} Acknowledgment}
This work was supported by the U.S. DOE, Office of Science, Basic Energy Sciences, Materials Science and Engineering Division under contract No. DE-SC0022060


\providecommand{\noopsort}[1]{}\providecommand{\singleletter}[1]{#1}%

\end{document}